\documentstyle[aps,preprint,eqsecnum,floats,epsf]{revtex}
\tighten
 \draft
\begin{document}

\preprint{}
\title{Generalized Jaynes-Cummings Model with 
Intensity-Dependent and Non-resonant Coupling}
\author{A.~N.~F. Aleixo\thanks{Electronic address:
        {\tt aleixo@nucth.physics.wisc.edu}}}
\address{Department of Physics, University of Wisconsin\\
         Madison, Wisconsin 53706 USA\thanks{{\tt Permanent address}:
Instituto de F\'{\i}sica, Universidade Federal 
         do Rio de Janeiro, RJ - 
         Brazil.}}
\author{A.~B. Balantekin\thanks{Electronic address:
        {\tt baha@nucth.physics.wisc.edu}},}
\address{Max-Planck-Institut f\"ur Kernphysik,
Postfach 103980, D-69029 Heidelberg, Germany
\thanks{{\tt Permanent address:}Department of Physics, University of
Wisconsin,
         Madison, Wisconsin 53706 USA}}
\author{M.~A. C\^andido Ribeiro\thanks{Electronic address:
         {\tt macr@df.ibilce.unesp.br}},}
\address{Departamento de F\'{\i}sica - Instituto de Bioci\^encias,
         Letras e Ci\^encias Exatas\\
         UNESP, S\~ao Jos\'e do Rio Preto, SP - Brazil}
\date{\today}
\maketitle

\begin{abstract}
  We study the intensity-dependent and nonresonant  Jaynes-Cummings
  Hamiltonian when the field is described by an arbitrary
  shape-invariant system.  We determine the eigenstates, eigenvalues,
  time evolution matrix and the population inversion  matrix factor. 
\end{abstract}


\newpage
\section{Introduction}

In the preceding paper \cite{ref17a} we extended our earlier work
\cite{ref17} describing interactions between a two-level system and a
shape-invariant \cite{ref3,ref5,ref6} system. Here we present another
generalization. 

The model studied in Ref. \cite{ref17} is a generalization of the
Jaynes-Cummings model \cite{ref18}. In the standard Jaynes-Cummings
model the ``field'' is described by a harmonic oscillator. In our
generalization it can be described by any shape-invariant system. In
developing this model we made extensive use of the algebraic approach
\cite{ref5,ref6} to the supersymmetric quantum mechanics
\cite{ref2}. In this paper we further generalize the model to one with
intensity-dependent interactions. 

The standard Jaynes-Cummings model is an idealized model describing
the interaction of matter with electromagnetic radiation. A variant of
the  Jaynes-Cummings model takes the coupling between matter and the
radiation to depend on the intensity of the electromagnetic field
\cite{ref24,ref25,ref26,ref26a}. This model has great relevance  since
this kind of interaction means that the coupling is proportional to
the amplitude of the field which is a very simple case of a nonlinear
interaction corresponding to a more realistic physical situation. The
results of this model can also give insight into the behavior of other
quantum systems with strong nonlinear interactions. 

\section{The Generalized Intensity-Dependent and Non-resonant 
Jaynes-Cummings Hamiltonian}

The expression of the intensity-dependent and
non-resonant Jaynes-Cummings Hamiltonian can be written as 
\begin{equation}
\hat {\bf H} = \hat A^\dagger\hat A + {1\over 2}\left[\hat A,\hat
A^\dagger\right]\left(\hat\sigma_3+1\right) + \alpha\left(\hat\sigma_
+ \hat A\;\sqrt{\hat A^\dagger\hat A}\, + \hat\sigma_-\sqrt{\hat 
A^\dagger\hat
A}\;\hat A^\dagger\right) + \hbar\Delta\,\hat\sigma_3\,,
\label{eqjcnr}
\end{equation}
where $\alpha$ is a constant related with the coupling strength,
$\Delta$ is a constant related with the detuning of the system  and
$\hat\sigma_i$, with $i=1,\,2,\, {\rm and}\,\, 3$, are the Pauli 
matrices. 

However, the harmonic oscillator systems, used in this context, is
only the simplest example of supersymmetric and shape-invariant
potential. Our goal at this point is to generalize that Hamiltonian
for all supersymmetric and shape-invariant systems. With this purpose
we introduce the operators
\begin{mathletters}
\label{eqso}
\begin{eqnarray}
& & \hat {\bf S} = \hat\sigma_+\hat A + \hat\sigma_-\hat A^\dagger\\ 
& & \hat
{\bf S}_i = \hat\sigma_+\hat A\;\sqrt{\hat A^\dagger\hat A}\, +
\hat\sigma_-\sqrt{\hat A^\dagger\hat A}\;\hat A^\dagger\,,
\end{eqnarray}
\end{mathletters}
where 
\begin{equation}
\hat\sigma_\pm = {1\over 2}\left( \hat\sigma_1\pm i\hat\sigma_2\right)\,,
\label{eqsig}
\end{equation}
and, now, the operators $\hat A$ and $\hat A^\dagger$ satisfy the
shape invariance condition \cite{ref17}. Using this definition
we can decompose the Jaynes-Cummings Hamiltonian in the form
\begin{equation}
\hat {\bf H} = \hat {\bf H}_o + \hat {\bf H}_{int}\,,
\label{eqjcnr1}
\end{equation}
where 
\begin{mathletters}
\label{eqhint}
\begin{eqnarray}
& & \hat {\bf H}_o = \hat {\bf S}^2\,,\\ & & \hat {\bf H}_{int} =
\alpha\,\hat {\bf S}_i + \hbar\Delta\,\hat\sigma_3\,. 
\end{eqnarray}
\end{mathletters}
We search for the eigenstates of $\hat {\bf H}$ and, in this case, it
is more convenient to work with its $B$-operator expressions, which
can be written as \cite{ref17} 
\begin{mathletters}
\label{eqso2}
\begin{eqnarray}
& & \hat {\bf S}^2 = \left[ \matrix{\hat T\hat B_-\hat B_+\hat
T^\dagger  & 0\cr 0 & \hat B_+\hat B_- \cr}\right] \equiv  \left[
\matrix{\hat H_2 & 0\cr 0 & \hat H_1 \cr}\right]\\ & & \hat {\bf
H}_{int} = \alpha \left[ \matrix{\beta & \hat T\hat B_-\;\sqrt{\hat
B_+\hat B_-}\cr \sqrt{\hat B_+\hat B_-}\;\hat B_+ \hat T^\dagger &
-\beta\cr}\right]\; = \;\alpha \left[ \matrix{\beta & \hat T\hat
B_-\;\sqrt{\hat H_1}\cr \sqrt{\hat H_1}\;\hat B_+ \hat T^\dagger &
-\beta\cr}\right]\,,
\end{eqnarray}
\end{mathletters} 
where \ $\beta = \hbar\Delta/\alpha$.  We use the same notation as the
preceding paper \cite{ref17a}.  There we show that the states 
\begin{equation}
\mid \Psi_m^{(\pm)}\rangle =  \left[ \matrix{\hat T & 0\cr 0 & \pm 1
\cr}\right]  \left[ \matrix{C_m^{(\pm)}\mid m\rangle\cr
C_{m+1}^{(\pm)}\mid m+1\rangle \cr}\right], \qquad m=0,1,2, \cdots 
\label{eqpsm+-}
\end{equation}
are the eigenstates of the operator $\hat {\bf S}^2$
\begin{equation}
\hat {\bf S}^2\mid \Psi_m^{(\pm)}\rangle = {\cal E}_{m+1}\mid
\Psi_m^{(\pm)}\rangle\,. 
\label{eqs2psm}
\end{equation}
In this case \ $C_{m,m+1}^{(\pm)} \equiv C_{m,m+1}^{(\pm)}
\left[R(a_1), R(a_2), R(a_3), \dots\right]$ \ are auxiliary
coefficients and, $\mid m\rangle$ and  $\mid m+1\rangle$ are the
abbreviated notation for the states $\mid \psi_m\rangle$ and $\mid
\psi_{m+1}\rangle$ \cite{ref17a}. 

At this point, we observe that the wave-state orthonormalization
conditions imply in the following relations among the $C$'s real
coefficients
\begin{mathletters}
\label{eqcpm}
\begin{eqnarray}
\langle\Psi_m^{(\pm)}\mid \Psi_m^{(\pm)}\rangle &=& \left[
C_m^{(\pm)}\right]^2 + \left[ C_{m+1}^{(\pm)}\right]^2 = 1 \\
\langle\Psi_m^{(\mp)}\mid \Psi_m^{(\pm)}\rangle &=&
C_m^{(\pm)}C_m^{(\mp)} - C_{m+1}^{(\pm)}C_{m+1}^{(\mp)} = 0 \,. 
\end{eqnarray}
\end{mathletters}
Now, considering that $\hat {\bf S}^2$ and $\hat {\bf H}_{int}$
commute then it is possible to find a common set of eigenstates. We
can use this fact to determine the eigenvalues of  $\hat {\bf
H}_{int}$ and the relations among the $C$'s coefficients. For that we
need to calculate
\begin{equation}
\hat {\bf H}_{int}\mid \Psi_m^{(\pm)}\rangle =  \lambda_m^{(\pm)}\mid
\Psi_m^{(\pm)}\rangle\,,
\label{eqhipsi}
\end{equation}
where $\lambda_m^{(\pm)}$ are the eigenvalues to be determined.  Using
the Eqs.~(\ref{eqso}), (\ref{eqhint}) and (\ref{eqpsm+-}), the last
eigenvalue equation can be rewritten in a matrix form as 
\begin{equation}
\alpha\left[ \matrix{\beta & \hat T\hat B_-\;\sqrt{\hat H_1} \cr
\sqrt{\hat H_1}\;\hat B_+\hat T^\dagger & -\beta \cr}\right] \left[
\matrix{\hat T & 0\cr 0 & \pm 1\cr}\right]\left[
\matrix{C_m^{(\pm)}\mid m\rangle\cr  C_{m+1}^{(\pm)}\mid
m+1\rangle}\right] =  \lambda_m^{(\pm)}\left[ \matrix{C_m^{(\pm)}\mid
m\rangle\cr  C_{m+1}^{(\pm)}\mid m+1\rangle}\right] \,,
\label{eqhimt}
\end{equation}
Since the $C$'s coefficients commute with the $\hat A$ or $\hat
A^\dagger$ operators, then the last matrix equation permits to obtain
the following equations
\begin{mathletters}
\label{eqmat1}
\begin{eqnarray}
& &\left[ \alpha\beta-\lambda_m^{(\pm)}\right]\left(\hat T
C_m^{(\pm)}\hat T^\dagger\right)\hat T\mid m\rangle \pm
\alpha\,C_{m+1}^{(\pm)}\hat T\hat B_-\;\sqrt{\hat H_1} 
\mid m+1\rangle= 0 \\ & &\alpha\left( \hat T C_m^{(\pm)}
\hat T^\dagger\right)
\sqrt{\hat H_1}\;\hat B_+\mid m\rangle \mp \left[ \alpha\beta +
\lambda_m^{(\pm)}\right]C_{m+1}^{(\pm)} \mid m+1\rangle = 0 \,. 
\end{eqnarray}
\end{mathletters}

Introducing the operator \cite{ref7}
\begin{equation}
\hat Q^\dagger = \left(\hat B_+\hat B_-\right)^{-1/2}\hat B_+ =
\left(\hat H_1\right)^{-1/2}\hat B_+
\label{eqq+}
\end{equation}
one can write the normalized eigenstate of $\hat H_1$ as 
\begin{equation}
\mid m\rangle = \left( \hat Q^\dagger\right)^m\mid 0\rangle\,,
\label{eqsmq}
\end{equation}
and, with Eqs.~(\ref{eqq+}) and (\ref{eqsmq}) we can show  that
\cite{ref17}
\begin{mathletters}
\label{eqbm}
\begin{eqnarray}
& & \hat B_+\mid m\rangle = \sqrt{{\cal E}_{m+1}}\mid m+1\rangle\,,\\
& &\hat T\hat B_-\mid m+1\rangle = \sqrt{{\cal E}_{m+1}}\,\hat T\mid
m\rangle\,. 
\end{eqnarray}
\end{mathletters}
Therefore, we have that 
\begin{eqnarray}
\hat T\hat B_-\;\sqrt{\hat H_1}\mid m+1\rangle &=&   \hat T\hat
B_-\;\sqrt{{\cal E}_{m+1}}\mid m+1\rangle\nonumber\\ &=& \sqrt{{\cal
E}_{m+1}}\;\hat T\hat B_-\mid m+1\rangle\nonumber\\ &=& {\cal
E}_{m+1}\;\hat T\mid m\rangle\,,
\label{eqtbh1}
\end{eqnarray}
and 
\begin{eqnarray}
\sqrt{\hat H_1}\;\hat B_+\mid m\rangle &=&   \sqrt{\hat
H_1}\;\sqrt{{\cal E}_{m+1}}\mid m+1\rangle\nonumber\\ &=& \sqrt{{\cal
E}_{m+1}}\;\sqrt{\hat H_1}\mid m+1\rangle\nonumber\\ &=& {\cal
E}_{m+1}\mid m+1\rangle\,,
\label{eqh1b+}
\end{eqnarray}
Using Eqs.~(\ref{eqtbh1}) and (\ref{eqh1b+}), then Eqs.~(\ref{eqmat1})
take the form 
\begin{mathletters}
\label{eqmat2}
\begin{eqnarray}
& &\left\{\left[ \alpha\beta-\lambda_m^{(\pm)}\right]\left(\hat T
C_m^{(\pm)}\hat T^\dagger\right) \pm \alpha \;{\cal
E}_{m+1}\,C_{m+1}^{(\pm)}\right\}\hat T\mid m\rangle = 0 \\ &
&\left\{\alpha \;{\cal E}_{m+1}\left( \hat T C_m^{(\pm)}\hat
T^\dagger\right) \mp \left[ \alpha\beta +
\lambda_m^{(\pm)}\right]C_{m+1}^{(\pm)}\right\} \mid m+1\rangle = 0
\,. 
\end{eqnarray}
\end{mathletters}
From Eqs.~(\ref{eqmat2}) it follows that 
\begin{equation}
\lambda_m^{(\pm)} = \pm\alpha\sqrt{{\cal E}_{m+1}^2 + \beta^2}\,,
\label{eqlb}
\end{equation}
and 
\begin{equation}
C_{m+1}^{(\pm)} = \left( {\sqrt{{\cal E}_{m+1}^2 + \beta^2}\mp
\beta\over {\cal E}_{m+1}}\right)\, \left(\hat T C_m^{(\pm)}\hat
T^\dagger \right)\,. 
\label{eqccm}
\end{equation}
Eqs.~(\ref{eqcpm}) and (\ref{eqccm}) imply that 
\begin{equation}
C_{m+1}^{(\pm)} = C_m^{(\mp)}\,,
\label{eqccm1}
\end{equation}
and the eigenstates and eigenvalues of the generalized
intensity-dependent and non-resonant Jaynes-Cummings Hamiltonian can be
written as 
\begin{equation}
E_m^{(\pm)} = {\cal E}_{m+1} \pm \sqrt{\alpha^2\; {\cal E}_{m+1}^2 +
\hbar^2\Delta^2}\,,
\label{eqemjc}
\end{equation}
and 
\begin{equation}
\mid \Psi_m^{(\pm)}\rangle =  \left[ \matrix{\hat T & 0\cr 0 & \pm 1
\cr}\right]  \left[ \matrix{C_m^{(\pm)}\mid m\rangle\cr
C_m^{(\mp)}\mid m+1\rangle \cr}\right], \qquad m=0,1,2, \cdots 
\label{eqesjc}
\end{equation}

{\bf a) The Intensity-Dependent Resonant Limit}
\bigskip

From these general results we can verify two simple limiting
cases. The first one corresponds to the resonant situation, which is
for \ $\Delta = 0$ \ $(\beta = 0)$. Using these conditions into
Eqs.~({\ref{eqccm}) and (\ref{eqemjc}) and Eqs.~(\ref{eqcpm}) we can
promptly conclude that
\begin{equation}
E_m^{(\pm)} = \left( 1 \pm \alpha\right)\,{\cal E}_{m+1}\,,
\label{eqemjcr}
\end{equation}
and 
\begin{equation}
C_{m+1}^{(\pm)} = \hat T C_m^{(\pm)}\hat T^\dagger = C_m^{(\pm)} =
{1\over \sqrt{2}}\,. 
\label{eqccmr}
\end{equation}
Therefore the intensity-dependent Jaynes-Cummings resonant 
eigenstate is given by
\begin{equation}
\mid \Psi_m^{(\pm)}\rangle =  {1\over \sqrt{2}} \left[ \matrix{\hat T
& 0\cr 0 & \pm 1 \cr}\right]  \left[ \matrix{ \mid m\rangle\cr \mid
m+1\rangle \cr}\right], \qquad m=0,1,2, \cdots 
\label{eqesjcr}
\end{equation}
If we compare this last particular result with that one found in the
reference \cite{ref17}, we conclude that the intensity-dependent and
intensity-independent generalized Jaynes-Cummings Hamiltonians have
the same eigenstates in the resonant situation. 

\bigskip

{\bf b) The Standard Intensity-Dependent Jaynes-Cummings Limit}
\bigskip

The second important limit corresponds to the standard
intensity-dependent Jaynes-Cummings case, related with the harmonic
oscillator system. In this limit we have that  \ $\hat T = \hat
T^\dagger \longrightarrow 1$, \ $\hat B_- \longrightarrow \hat a$, \
$\hat B_+ \longrightarrow \hat a^\dagger$, \ $\Delta = \omega -
\omega_o$ \ and ${\cal E}_{m+1} = (m+1)\hbar\omega$. \ Using these
conditions in  Eqs.~({\ref{eqccm}), (\ref{eqemjc}) and
Eqs.~(\ref{eqcpm}) we  can promptly conclude that 
\begin{equation}
E_m^{(\pm)} = (m+1)\hbar\omega \pm \hbar\sqrt{\alpha^2\omega^2 
(m+1)^2 +  (\omega-\omega_o)^2}\,,
\label{eqemjcs}
\end{equation}
and  
\begin{equation}
C_{m+1}^{(\pm)} = \gamma_m^{(\pm)}C_m^{(\pm)} = C_m^{(\mp)} =  
{1\over \sqrt{1 + \left(\gamma_m^{(\mp)}\right)^2}}\,,
\label{eqccms}
\end{equation}
where 
\begin{mathletters}
\label{eqgd}
\begin{eqnarray}
& & \gamma_m^{(\pm)} = \sqrt{1 + \delta_m^2} \mp \delta_m\,,\\ & &
\delta_m = {\omega-\omega_o\over \alpha\omega (m+1)}\,. 
\end{eqnarray}
\end{mathletters}
Therefore the standard intensity-dependent Jaynes-Cummings 
eigenstate, written in a matrix form, is given by 
\begin{equation}
\mid \Psi_m^{(\pm)}\rangle =  {1\over \sqrt{1 +
\left(\gamma_m^{(\pm)}\right)^2}}  \left[ \matrix{ 1 & 0\cr 0 & \pm
\gamma_m^{(\pm)} \cr}\right]  \left[ \matrix{ \mid m\rangle\cr \mid
m+1\rangle \cr}\right], \qquad m=0,1,2, \cdots\,. 
\label{eqesjcs}
\end{equation}

\section{Time Evolution of the System}

To resolve the the time-dependent Schr\"odinger equation for
intensity-dependent and non-resonant Jaynes-Cummings systems  
\begin{equation}
i\hbar {\partial \over \partial t} \mid \Psi (t)\rangle =  \left(\hat
{\bf H}_o + \hat {\bf H}_{int}\right)\mid \Psi (t)\rangle
\label{eqschr}
\end{equation} 
we can write the state as 
\begin{equation}
\mid \Psi (t)\rangle = \exp{\left(-i\hat {\bf
H}_ot/\hbar\right)}\,\mid \Psi_i(t)\rangle\,,
\label{eqpsint}
\end{equation}
and, by substituting this into Schr\"odinger equation and taking into
account the commutation property between $\hat {\bf H}_o$ and $\hat
{\bf H}_{int}$, we obtain 
\begin{equation}
i\hbar {\partial \over \partial t} \mid \Psi_i(t)\rangle =  \hat {\bf
H}_{int}\,\mid \Psi_i(t)\rangle\,,
\label{eqschrint}
\end{equation} 
Now, we can introduce the evolution matrix $\hat {\bf U}_i(t,0)$, 
related with the interaction Hamiltonian, by 
\begin{equation}
\mid \Psi_i(t)\rangle = \hat {\bf U}_i(t,0)\,\mid \Psi_i(0)\rangle\,. 
\label{eqevu}
\end{equation}
with 
\begin{equation}
i\hbar {\partial \over \partial t} \hat {\bf U}_i(t,0) =  \hat {\bf
H}_{int}\,\hat {\bf U}_i(t,0)\,,
\label{eqschri}
\end{equation} 
that is, in matrix form, written as
\begin{equation}
i\hbar\left[ \matrix{\hat U_{11}^\prime & \hat U_{12}^\prime \cr  \hat
U_{21}^\prime & \hat U_{22}^\prime \cr}\right] =  \alpha\left[
\matrix{\beta & \hat T\hat B_-\;\sqrt{\hat H_1} \cr \sqrt{\hat
H_1}\;\hat B_+\hat T^\dagger & -\beta \cr}\right]  \left[ \matrix{\hat
U_{11} & \hat U_{12} \cr \hat U_{21} & \hat U_{22} \cr}\right]\,,
\label{eqdevo}
\end{equation}
where the primes denote the time derivative.  One fast way to
diagonalize the evolution matrix differential equation is by
differentiating Eq.~({\ref{eqschri}) with respect to time. After that,
if we use again the same Eq.~({\ref{eqschri}), we find 
\begin{equation}
i\hbar {\partial^2 \over \partial t^2} \hat {\bf U}_i(t,0) =  \hat
{\bf H}_{int}\, {\partial \over \partial t}\hat {\bf U}_i(t,0) =
{1\over i\hbar}\hat {\bf H}_{int}^2 \hat {\bf U}_i(t,0)\,,
\label{eqschr3}
\end{equation} 
which can be written as 
\begin{equation}
\left[ \matrix{\hat U_{11}^{\prime\prime} & \hat U_{12}^{\prime\prime}
\cr \hat U_{21}^{\prime\prime} & \hat U_{22}^{\prime\prime}
\cr}\right] =  -\left[ \matrix{\hat \omega_1 & 0 \cr  0 & \hat
\omega_2 \cr}\right]  \left[ \matrix{\hat U_{11} & \hat U_{12} \cr
\hat U_{21} & \hat U_{22} \cr}\right]\,,
\label{eqdevo1}
\end{equation}
where 
\begin{mathletters}
\label{eqfreq}
\begin{eqnarray}
& & \hbar\hat \omega_1 = \alpha\sqrt{(\hat T\hat B_-\hat B_+\hat
T^\dagger)^2  + \beta^2} = \sqrt{\alpha^2\,\hat H_2^2 +
(\hbar\Delta)^2}\,,\\ & & \hbar\hat \omega_2 = \alpha\sqrt{(\hat
B_+\hat B_-)^2 + \beta^2} = \sqrt{\alpha^2\,\hat H_1^2 +
(\hbar\Delta)^2}\,. 
\end{eqnarray}
\end{mathletters}
Now, since by initial conditions \  $\hat {\bf U}_i(0,0) = \hat {\bf
I}$, \ then we can write the solution of the evolution matrix
differential equation ({\ref{eqschr3}) as 
\begin{equation}
\hat {\bf U}_i(t,0) = \left[ \matrix{\cos{(\hat\omega_1 t)} &
\sin{(\hat\omega_1 t)}\,\hat C\cr \sin{(\hat\omega_2 t)}\,\hat D &
\cos{(\hat\omega_2 t)} \cr}\right]\,,
\label{eqevo3}
\end{equation}
and the $\hat C$ and $\hat D$ operators can be determined by the
unitary transformation conditions 
\begin{equation}
\hat {\bf U}_i^\dagger (t,0)\,\hat {\bf U}_i (t,0) =  \hat {\bf U}_i
(t,0)\,\hat {\bf U}_i^\dagger (t,0) = \hat {\bf I}\,. 
\label{equni}
\end{equation}
Following the same steps used in the appendix A of the reference
\cite{ref17a} we can conclude that to satisfy the unitary conditions
(\ref{equni}) these operators must have the form 
\begin{mathletters}
\label{eqcdfin}
\begin{eqnarray}
& & \hat C = -\hat D^\dagger = {i\over \hat H_2^{1/4}}\; \sqrt{\hat
T\hat B_-}\\ & & \hat D = -\hat C^\dagger = \sqrt{\hat B_+\hat
T^\dagger}\; {i\over \hat H_2^{1/4}}\,. 
\end{eqnarray}
\end{mathletters}
Therefore, we can write the final expression of the time evolution
matrix of the system as 
\begin{equation}
\hat {\bf U}_i(t,0) = \left[ \matrix{\cos{(\hat\omega_1 t)} &
\sin{(\hat\omega_1 t)}\,\hat C\cr -\sin{(\hat\omega_2 t)}\,\hat
C^\dagger &  \cos{(\hat\omega_2 t)} \cr}\right]\,. 
\label{eqevo4}
\end{equation}

For Jaynes-Cummings systems an important physical quantity to see how
the system under consideration evolves in time is the population
inversion factor \cite{ref25,ref21,ref23}, defined by 
\begin{equation}
\hat {\bf W}(t) \equiv \hat\sigma_+(t)\;\hat\sigma_-(t) -
\hat\sigma_-(t)\;\hat\sigma_+(t) = \hat\sigma_3(t)\,,
\label{eqinpo}
\end{equation}
where the time dependence of the operators is related with the
Heisenberg picture. In this case, the time evolution of the population
inversion factor will be given by 
\begin{equation}
{d\hat \sigma_3(t)\over dt} = {1\over i\hbar}\hat {\bf U}_i^\dagger
(t,0)\left[\hat\sigma_3, \hat{\bf H}\right]\hat {\bf U}_i(t,0)\,,
\label{eqds3}
\end{equation}
and since we have 
\begin{equation}
\left[ \hat\sigma_3, \hat {\bf H}\right] = \alpha\left[ \hat\sigma_3,
\hat {\bf S}_i\right] = -2\alpha\,\hat {\bf S}_i \,\hat\sigma_3\,,
\label{eqcs3h}
\end{equation}
then Eq.~(\ref{eqds3}) can be written as 
\begin{equation}
{d\hat \sigma_3(t)\over dt} = {2i\alpha\over\hbar}\;\hat {\bf
S}_i(t)\, \hat \sigma_3(t)\,. 
\label{eqds4}
\end{equation}
We can obtain a differential equation with constant coefficients for
$\hat\sigma_3(t)$ by taking the time derivative of Eq.~(\ref{eqds4}) 
\begin{equation}
{d^2\hat \sigma_3(t)\over dt^2} = {2i\alpha\over\hbar}\; \left\{
{d\hat {\bf S}_i(t)\over dt}\,\hat \sigma_3(t) + \hat {\bf S}_i(t)\,
{d\hat \sigma_3(t)\over dt}\right\}\,. 
\label{eqds5}
\end{equation}
Having in mind that 
\begin{equation}
{d\hat {\bf S}_i(t)\over dt} = {1\over i\hbar}\hat {\bf U}_i^\dagger
(t,0)\left[\hat {\bf S}_i, \hat{\bf H}\right]\hat {\bf U}_i(t,0)\,,
\label{eqds}
\end{equation}
and,  
\begin{equation}
\left[ \hat {\bf S}_i, \hat {\bf H}\right] = \alpha\beta\left[ \hat
{\bf S}_i, \hat\sigma_3\right] = 2\alpha\beta\,\hat {\bf S}_i\,
\hat\sigma_3\,,
\label{eqcshh}
\end{equation}
we can conclude that 
\begin{equation}
{d\hat {\bf S}_i(t)\over dt} = -{2i\alpha\beta\over\hbar}\;\hat {\bf
S}_i(t)\, \hat \sigma_3(t)\,. 
\label{eqdss}
\end{equation}
Now using Eqs.~(\ref{eqds4}) and (\ref{eqdss}) into Eq.~(\ref{eqds5})
we obtain 
\begin{equation}
{d^2\hat \sigma_3(t)\over dt^2} + \hat {\bf \Theta}^2\, \hat
\sigma_3(t) = \hat {\bf F}(t) 
\label{eqds6}
\end{equation}
where 
\begin{mathletters}
\label{eqomft}
\begin{eqnarray}
& & \hat {\bf \Theta}^2 = {4\alpha^2\over\hbar^2}\,\hat {\bf S}_i^2\\
& & \hat {\bf F}(t) = {4\alpha^2\beta\over\hbar^2}\, \hat {\bf
U}_i^\dagger (t,0)\,\hat {\bf S}_i\,\hat {\bf U}_i(t,0)\,. 
\end{eqnarray}
\end{mathletters}

The Eq.~(\ref{eqdss}) corresponds to a non-homogeneous linear
differential equation for $\hat \sigma_3(t)$ with constant
coefficients since  $\hat {\bf S}_i^2$ and $\hat {\bf H}$ commute and,
therefore, $\hat {\bf \Theta }$ is a constant of the motion. The
general solution of this differential equation can be written as  
\begin{equation}
\hat \sigma_3(t) = \hat\sigma^H(t) + \hat\sigma^P(t)\,,
\label{eqs3gen}
\end{equation}
and each matrix element of the homogeneous solution, satisfies the
differential equation 
\begin{equation}
{d^2\hat\sigma^H_{jk}(t)\over dt^2} +
\hat\nu_j^2\,\hat\sigma^H_{jk}(t) = 0\,,  \qquad j,\,k = 1, {\rm or}\;
2\,,
\label{eqs3ho}
\end{equation}
with 
\begin{mathletters}
\label{eqfreqo}
\begin{eqnarray}
& & \hbar\hat \nu_1 = 2\alpha\,\hat T\hat B_-\hat B_+\hat T^\dagger =
2\alpha\,\hat H_2\,,\\ & & \hbar\hat \nu_2 = 2\alpha\,\hat B_+\hat B_-
= 2\alpha\,\hat H_1\,. 
\end{eqnarray}
\end{mathletters}
The solution of Eq.~(\ref{eqs3ho}) is given by 
\begin{equation}
\hat\sigma^H_{jk}(t) = \hat y_j(t)\;\hat c_{jk} + \hat z_j(t)\;\hat
d_{jk}\,,
\label{eqs3hog}
\end{equation}
where 
\begin{mathletters}
\label{eqlisol}
\begin{eqnarray}
& & \hat y_j(t) = \cos{(\hat\nu_jt)}\\ & & \hat z_j(t) =
\sin{(\hat\nu_jt)}\,,
\end{eqnarray}
\end{mathletters}
and the coefficients $\hat c_{jk}$ and $\hat d_{jk}$ can be determined
by the initial conditions. 

The matrix elements of the particular solution of the
$\hat\sigma_3(t)$ differential equation needs to satisfy 
\begin{equation}
{d^2\hat\sigma^P_{jk}(t)\over dt^2} +
\hat\nu_j^2\,\hat\sigma^P_{jk}(t) = \hat F_{jk}(t)\,,  \qquad j,\,k =
1, {\rm or}\; 2\,,
\label{eqs3pt}
\end{equation}
and can be obtained by the variation of parameter or by Green function
methods, giving 
\begin{equation}
\hat\sigma_{jk}^P(t) = \hat\nu_j^{-1}\,\left\{ \hat z_j(t) \int_0^t
\xi\,\hat y_j(\xi)\,\hat F_{jk}(\xi) -  \hat y_j(t)\int_0^t
d\xi\,\hat z_j(\xi)\,\hat F_{jk}(\xi) \right\}\,,
\label{eqvarp}
\end{equation}
where we used that the Wronskian of the system of solutions $\hat
y_j(t)$ and $\hat z_j(t)$ is given by $\hat\nu_j$. 

After we determine the elements of the $\hat {\bf F}(t)$-matrix,
it is necessary to resolve the integrals in Eq.~(\ref{eqvarp}) to
obtain the explicit expression of the particular solution.  In the
appendix we show that, using Eqs.~(\ref{eqso}), (\ref{eqevo4}), and
(\ref{eqomft}), it is possible to conclude that these matrix elements
can be written as 
\begin{mathletters}
\label{eqs3p11}
\begin{eqnarray}
\hat\sigma_{11}^P(t) &=& i{\gamma\over 2}\hat\nu_1^{-1}\sqrt{\hat
T\hat B_-}\left\{\hat z_2(t)\,{\cal
G}_{CS}^{(+)}(t;\hat\nu_2,\hat\omega_2,\hat\omega_1) -  \hat
y_2(t)\,{\cal G}_{SS}^{(+)}(t;\hat\nu_2,\hat\omega_2, \hat\omega_1)
\right\}\hat H_2^{3/4}\nonumber\\ &+& i{\gamma\over
2}\hat\nu_1^{-1}\hat H_2^{3/4}\left\{\hat z_1(t)\, {\cal
G}_{SC}^{(-)}(t;\hat\nu_1,\hat\omega_1,\hat\omega_2) -  \hat
y_1(t)\,{\cal G}_{CC}^{(-)}(t;\hat\nu_1,\hat\omega_1, \hat\omega_2)
\right\}\sqrt{\hat B_+\hat T^\dagger}\,,\\
\hat\sigma_{12}^P(t) &=& {\gamma\over 2}\hat\nu_1^{-1}\sqrt{\hat T\hat
B_-}\left\{\hat z_2(t)\,{\cal
G}_{CC}^{(+)}(t;\hat\nu_2,\hat\omega_2,\hat\omega_1) -  \hat
y_2(t)\,{\cal G}_{SC}^{(+)}(t;\hat\nu_2,\hat\omega_2, \hat\omega_1)
\right\}\sqrt{\hat H_2\hat T\hat B_-}\nonumber\\ &+& {\gamma\over
2}\hat\nu_1^{-1}\hat H_2^{3/4}\left\{\hat z_1(t)\, {\cal
G}_{SS}^{(-)}(t;\hat\nu_1,\hat\omega_1,\hat\omega_2) +  \hat
y_1(t)\,{\cal G}_{CS}^{(-)}(t;\hat\nu_1,\hat\omega_1, \hat\omega_2)
\right\}\hat H_1^{1/4}\,,\\
\hat\sigma_{21}^P(t) &=& {\gamma\over 2}\hat\nu_2^{-1}\sqrt{\hat
B_+\hat T^\dagger\hat H_2}\left\{\hat z_1(t)\,{\cal
G}_{CC}^{(+)}(t;\hat\nu_1,\hat\omega_1,\hat\omega_2) -  \hat
y_1(t)\,{\cal G}_{SC}^{(+)}(t;\hat\nu_1,\hat\omega_1, \hat\omega_2)
\right\}\sqrt{\hat B_+\hat T^\dagger}\nonumber\\ &+& {\gamma\over
2}\hat\nu_2^{-1}\hat H_1^{1/4}\left\{\hat z_2(t)\, {\cal
G}_{SS}^{(-)}(t;\hat\nu_2,\hat\omega_2,\hat\omega_1) -  \hat
y_2(t)\,{\cal G}_{CS}^{(-)}(t;\hat\nu_2,\hat\omega_2, \hat\omega_1)
\right\}\hat H_2^{3/4}\,,\\
\hat\sigma_{22}^P(t) &=& i{\gamma\over 2}\hat\nu_2^{-1}\sqrt{\hat
B_+\hat T^\dagger\hat H_2}\left\{\hat z_1(t)\,{\cal
G}_{CS}^{(+)}(t;\hat\nu_1,\hat\omega_1,\hat\omega_2) -  \hat
y_1(t)\,{\cal G}_{SS}^{(+)}(t;\hat\nu_1,\hat\omega_1, \hat\omega_2)
\right\}\hat H_1^{1/4}\nonumber\\ &+& i{\gamma\over
2}\hat\nu_2^{-1}\hat H_1^{1/4}\left\{\hat z_2(t)\, {\cal
G}_{SC}^{(-)}(t;\hat\nu_2,\hat\omega_2,\hat\omega_1) +  \hat
y_2(t)\,{\cal G}_{CC}^{(-)}(t;\hat\nu_2,\hat\omega_2, \hat\omega_1)
\right\}\sqrt{\hat H_2\hat T\hat B_-}\,,
\end{eqnarray}
\end{mathletters}
where \ $\gamma = 4\alpha^2\beta/\hbar^2$, \  and the auxiliary
functions are given by 
\begin{equation}
{\cal G}^{(\pm)}_{XY}(t;\hat p,\hat q,\hat r) = {\cal F}_{XY}(t;\hat
p-\hat q,\hat r) \pm {\cal F}_{XY}(t;\hat p+\hat q,\hat r)\,,\qquad X,
Y = C\;{\rm or}\;S\,,
\label{eqfgpm}
\end{equation}
with 
\begin{mathletters}
\label{eqfxy}
\begin{eqnarray}
{\cal F}_{CC}(t;\hat x,\hat w) &\equiv& \int_0^td\xi\,\cos{(\hat
x\xi)}\, \cos{(\hat w\xi)}\nonumber\\ &=& \sum_{m,n = 0}^\infty
(-1)^{m+n}{\hat x^{2m}\hat  w^{2n}\over (2m)!\,(2n)!}{t^{2m+2n+1}\over
(2m+2n+1)}\,\\
{\cal F}_{CS}(t;\hat x,\hat w) &\equiv& \int_0^td\xi\,\cos{(\hat
x\xi)}\, \sin{(\hat w\xi)}\nonumber\\ &=& \sum_{m,n = 0}^\infty
(-1)^{m+n}{\hat x^{2m}\hat  w^{2n+1}\over
(2m)!\,(2n+1)!}{t^{2m+2n+2}\over (2m+2n+2)}\,\\
{\cal F}_{SC}(t;\hat x,\hat w) &\equiv& \int_0^td\xi\,\sin{(\hat
x\xi)}\, \cos{(\hat w\xi)}\nonumber\\ &=& \sum_{m,n = 0}^\infty
(-1)^{m+n}{\hat x^{2m+1}\hat  w^{2n}\over
(2m+1)!\,(2n)!}{t^{2m+2n+2}\over (2m+2n+2)}\,\\
{\cal F}_{SS}(t;\hat x,\hat w) &\equiv& \int_0^td\xi\,\sin{(\hat
x\xi)}\, \sin{(\hat w\xi)}\nonumber\\ &=& \sum_{m,n = 0}^\infty
(-1)^{m+n}{\hat x^{2m+1}\hat  w^{2n+1}\over
(2m+1)!\,(2n+1)!}{t^{2m+2n+3}\over (2m+2n+3)}\,. 
\end{eqnarray}
\end{mathletters}
With these results for the particular solution we can conclude that 
\begin{equation}
\hat\sigma^P_{ij}(0) = 0 = {d\hat\sigma^P_{ij}(0)\over dt}\,. 
\label{eqsp0}
\end{equation}
Now, using Eqs.~(\ref{eqds4}), (\ref{eqs3gen}), (\ref{eqs3hog}) and
the initial conditions, we have 
\begin{mathletters}
\label{eqs3ini}
\begin{eqnarray}
& & \left[\hat\sigma_3(0)\right]_{ij} = \hat c_{ij}\\ & &
\left[{d\hat\sigma_3(0)\over dt}\right]_{ij} = {2i\alpha\over \hbar}\,
\left[\hat {\bf S}_i(0)\,\hat\sigma_3(0)\right]_{ij} = \hat\nu_i\,\hat
d_{ij}\,. 
\end{eqnarray}
\end{mathletters}
Therefore, the final expression for the elements of the population
inversion matrix of the system can be written as 
\begin{equation}
[\hat\sigma_3(t)]_{ij} = \cos{(\hat\nu_it)}\,
\left[\hat\sigma_3(0)\right]_{ij} +  {2i\alpha\over
\hbar}\,\sin{(\hat\nu_it)}\; \hat\nu_i^{-1} \left[\hat {\bf
S}(0)_i\,\hat\sigma_3(0)\right]_{ij} + \hat\sigma^P_{ij}(t)\,. 
\label{eqs3fin}
\end{equation}

Again, using these final results we can verify two limiting cases. 

\bigskip

{\bf a) The Intensity-Dependent Resonant Limit}
\bigskip

The first one corresponds to the intensity-dependent resonant $(\Delta
= 0)$. Eqs.~({\ref{eqfreq}), (\ref{eqevo4}), (\ref{eqfreqo}) and
(\ref{eqs3p11}) allow us to conclude that, in this case, the evolution
matrix of the system is given by 
\begin{equation}
\hat {\bf U}_i(t,0) = \left[ \matrix{\cos{\left({1\over
2}\hat\nu_1t\right)} &  \sin{\left({1\over 2}\hat\nu_1 t\right)}\,\hat
C\cr -\sin{\left({1\over 2}\hat\nu_2 t\right)}\, \hat C^\dagger &
\cos{\left({1\over 2}\hat\nu_2 t\right)} \cr}\right]\,. 
\label{eqevo5}
\end{equation}
and the elements of the population inversion of the system is 
\begin{equation}
[\hat\sigma_3(t)]_{ij} = \cos{(\hat\nu_it)}\,
\left[\hat\sigma_3(0)\right]_{ij} +  {2i\alpha\over
\hbar}\,\sin{(\hat\nu_it)}\; \hat\nu_i^{-1} \left[\hat {\bf
S}_i(0)\,\hat\sigma_3(0)\right]_{ij}\,. 
\label{eqs3fres}
\end{equation}

\bigskip

{\bf b) The Standard Intensity-Dependent Jaynes-Cummings Limit}
\bigskip

This second important limit corresponds to the case of the harmonic
oscillator system, and in this limit we have that  \ $\hat T = \hat
T^\dagger \longrightarrow 1$, \ $\hat B_- \longrightarrow \hat a$, \
$\hat B_+ \longrightarrow \hat a^\dagger$ \ and \ $[\hat a,\hat
a^\dagger] = \hbar\omega$. \  With these conditions the operators
$\hat\omega_1$  and $\hat\omega_2$ commute, and this fact permits to
evaluate the integrals related with the particular solution of the
population inversion elements using trigonometric product
relations. Using that and the expressions obtained in the appendix,
after a considerable amount of algebra and trigonometric product
relations we can show that is possible to write the expressions for
the $\hat\sigma_{ij}^P(t)$-matrix elements as 
\begin{mathletters}
\label{eqs3p11s}
\begin{eqnarray}
\hat\sigma_{11}^P(t) &=& i{\gamma\over 2}\hat\nu_1^{-1} \sqrt{\hat
a}\left\{{\cal K}_S(t;\hat\omega_2,\hat\omega_1,\hat\nu_2) - {\cal
K}_S(t;\hat\omega_2,-\hat\omega_1,\hat\nu_2) \right\} \left(\hat a\hat
a^\dagger\right)^{3/4}\nonumber\\ &-& i{\gamma\over
2}\hat\nu_1^{-1}\left(\hat a\hat a^\dagger\right)^{3/4} \left\{{\cal
K}_S(t;\hat\omega_2,\hat\omega_1,\hat\nu_1) - {\cal
K}_S(t;\hat\omega_2,-\hat\omega_1,\hat\nu_1) \right\} \sqrt{\hat
a^\dagger }\\
\hat\sigma_{12}^P(t) &=& {\gamma\over 2}\hat\nu_1^{-1}\sqrt{\hat a}
\left\{{\cal K}_C(t;\hat\omega_2,\hat\omega_1,\hat\nu_2) - {\cal
K}_C(t;\hat\omega_2,-\hat\omega_1,\hat\nu_2)\right\} \sqrt{\hat a\hat
a^\dagger\hat a}\nonumber\\ &-& {\gamma\over
2}\hat\nu_1^{-1}\left(\hat a\hat a^\dagger\right)^{3/4} \left\{{\cal
K}_C(t;\hat\omega_2,\hat\omega_1,\hat\nu_1) - {\cal
K}_C(t;\hat\omega_2,-\hat\omega_1,\hat\nu_1)\right\} \left(\hat
a^\dagger\hat a\right)^{1/4}\\
\hat\sigma_{21}^P(t) &=& {\gamma\over 2}\hat\nu_2^{-1} \sqrt{\hat
a^\dagger\hat a\hat a^\dagger} \left\{{\cal
K}_C(t;\hat\omega_2,\hat\omega_1,\hat\nu_1) + {\cal
K}_C(t;\hat\omega_2,-\hat\omega_1,\hat\nu_1)\right\} \sqrt{\hat
a^\dagger}\nonumber\\ &-& {\gamma\over 2}\hat\nu_2^{-1}\left(\hat
a^\dagger\hat a\right)^{1/4} \left\{{\cal
K}_C(t;\hat\omega_2,\hat\omega_1,\hat\nu_2) - {\cal
K}_C(t;\hat\omega_2,-\hat\omega_1,\hat\nu_2)\right\} \left(\hat a\hat
a^\dagger\right)^{3/4}\\
\hat\sigma_{22}^P(t) &=& i{\gamma\over 2}\hat\nu_2^{-1} \sqrt{\hat
a^\dagger\hat a\hat a^\dagger }\left\{{\cal
K}_S(t;\hat\omega_2,\hat\omega_1,\hat\nu_1)  + {\cal
K}_S(t;\hat\omega_2,-\hat\omega_1,\hat\nu_1)\right\} \left(\hat
a^\dagger\hat a\right)^{1/4}\nonumber\\ &-& i{\gamma\over
2}\hat\nu_2^{-1}\left(\hat a^\dagger\hat a\right)^{1/4} \left\{{\cal
K}_S(t;\hat\omega_2,\hat\omega_1,\hat\nu_2) + {\cal
K}_S(t;\hat\omega_2,-\hat\omega_1,\hat\nu_2) \right\}\sqrt{\hat a\hat
a^\dagger\hat a}\,,
\end{eqnarray}
\end{mathletters}
where, now, the auxiliary functions are given by 
\begin{mathletters}
\label{eqksc}
\begin{eqnarray}
{\cal K}_S(t;\hat p,\hat q,\hat r) &=& {\hat r\;\sin{\left [\left(\hat
p+\hat q\right)t\right ]}-\left(\hat p+\hat q\right)\,\sin{\left(\hat
rt\right)}\over \hat r^2-\left(\hat p+\hat q\right)^2}\\
{\cal K}_C(t;\hat p,\hat q,\hat r) &=& {\hat r\;\cos{\left [\left(\hat
p+\hat q\right)t\right ]}-\hat r\;\cos{\left(\hat rt\right)}\over \hat
r^2-\left(\hat p+\hat q\right)^2}\,. 
\end{eqnarray}
\end{mathletters}
Considering the expressions above we may easily verify that the
particular solution for the population inversion factor must still
satisfy the initial conditions (\ref{eqsp0}). Therefore, in this case
the final expression of the population inversion factor has the same
form given by Eq.~(\ref{eqs3fin}), with 
\begin{mathletters}
\label{eqfrosc}
\begin{eqnarray}
& & \hbar\hat \nu_1 = 2\alpha\;\hat a\hat a^\dagger
\,,\qquad\qquad\qquad \hbar\hat \nu_2 = 2\alpha\;\hat a^\dagger\hat
a\,,\\ & & \hbar\hat \omega_1 = \alpha\sqrt{(\hat a\hat a^\dagger)^2 +
\beta^2}\,,\qquad\qquad  \hbar\hat \omega_2 = \alpha\sqrt{(\hat
a^\dagger\hat a)^2 +\beta^2}\,. 
\end{eqnarray}
\end{mathletters}

\section{Conclusions}

In this article we introduced a class of shape-invariant bound-state
problems which represent two-level systems. The corresponding
coupled-channel Hamiltonians generalize the intensity-dependent and
non-resonant Jaynes-Cummings Hamiltonian.  These models are not only
interesting on their own account. Being exactly solvable
coupled-channels problems they may help to assess the validity and
accuracy of various approximate approaches to the coupled-channel
problems \cite{ref16}. 

\section*{ACKNOWLEDGMENTS}

This work was supported in part by the U.S. National Science
Foundation Grants No.\ PHY-9605140 and PHY-0070161 at the University
of Wisconsin, and in part by the University of Wisconsin Research
Committee with funds granted by the Wisconsin Alumni Research
Foundation.   A.B.B.\ acknowledges the support of the Alexander von
Humboldt-Stiftung. M.A.C.R.\  acknowledges the support of Funda\c
c\~ao de Amparo \`a Pesquisa do Estado de S\~ao Paulo (Contract No.\
98/13722-2). A.N.F.A.  acknowledges the support of Funda\c c\~ao
Coordena\c c\~ao de Aperfei\c coamento de Pessoal de N\'{\i}vel
Superior (Contract No.  BEX0610/96-8).  A.B.B.\ is grateful to the
Max-Planck-Institut f\"ur Kernphysik and H.A. Weidenm\"uller for the
very kind hospitality. 

\bigskip\bigskip\bigskip\bigskip\bigskip

\appendix{\bf Appendix}

\bigskip

In this appendix we show the steps necessary to obtain the explicit
expression of the elements of the population inversion particular
solution. To resolve the integrals in the Eq.~(\ref{eqvarp}), first we
need to determine the elements of the \ $\hat {\bf F}(t)$-matrix.  To
do that we can use Eqs.~(\ref{eqso}), (\ref{eqomft}),  and
(\ref{eqevo4}) to conclude that 
\begin{mathletters}
\label{eqlfelem}
\begin{eqnarray}
\hat F_{11}(t) &=& -\gamma\left\{\cos{(\hat\omega_1t)}\,\hat T\hat
B_-\,\sin{(\hat\omega_2t)}\,\hat C^\dagger + \hat
C\,\sin{(\hat\omega_2t)}\,\hat B_+\hat
T^\dagger\,\cos{(\hat\omega_1t)}\right\}\nonumber\\ &=&
i\gamma\left\{\sqrt{\hat T\hat
B_-}\,\cos{(\hat\omega_2t)}\,\sin{(\hat\omega_1t)}\,\hat H_2^{1/4} -
\hat
H_2^{1/4}\,\sin{(\hat\omega_1t)}\,\cos{(\hat\omega_2t)}\,\sqrt{\hat
B_+\hat T^\dagger}\right\}\\ \hat F_{12}(t) &=&
\gamma\left\{\cos{(\hat\omega_1t)}\,\hat T\hat
B_-\,\cos{(\hat\omega_2t)} - \hat C\,\sin{(\hat\omega_2t)}\,\hat
B_+\hat T^\dagger\,\sin{(\hat\omega_1t)}\,\hat C\right\}\nonumber\\
&=& \gamma\left\{\sqrt{\hat T\hat
B_-}\,\cos{(\hat\omega_2t)}\,\cos{(\hat\omega_1t)}\,\sqrt{\hat T\hat
B_-} + \hat H_2^{1/4}\,\sin{(\hat\omega_1t)}\,\sin{(\hat\omega_2t)}\,
\hat H_1^{1/4}\right\}\\ \hat F_{21}(t) &=&
\gamma\left\{\cos{(\hat\omega_2t)}\,\hat B_+\hat
T^\dagger\,\cos{(\hat\omega_1t)} - \hat
C^\dagger\,\sin{(\hat\omega_1t)}\,\hat T\hat
B_-\,\sin{(\hat\omega_2t)}\,\hat C^\dagger\right\}\nonumber\\ &=&
\gamma\left\{\sqrt{\hat B_+\hat
T^\dagger}\,\cos{(\hat\omega_1t)}\,\cos{(\hat\omega_2t)}\,\sqrt{\hat
B_+\hat T^\dagger} + \hat
H_1^{1/4}\,\sin{(\hat\omega_2t)}\,\sin{(\hat\omega_1t)}\, \hat
H_2^{1/4}\right\}\\ \hat F_{22}(t) &=& \gamma\left\{\hat
C^\dagger\,\sin{(\hat\omega_1t)}\, \hat T\hat
B_-\,\cos{(\hat\omega_2t)} + \cos{(\hat\omega_2t)}\,\hat B_+\hat
T^\dagger\,\sin{(\hat\omega_1t)}\,\hat C\right\}\nonumber\\ &=&
i\gamma\left\{\sqrt{\hat B_+\hat
T^\dagger}\,\cos{(\hat\omega_1t)}\,\sin{(\hat\omega_2t)}\,\hat
H_1^{1/4} -  \hat
H_1^{1/4}\,\sin{(\hat\omega_2t)}\,\cos{(\hat\omega_1t)}\,\sqrt{\hat
T\hat B_-}\right\}\,,
\end{eqnarray}
\end{mathletters}
where \ $\gamma = 4\alpha^2\beta/\hbar^2$, \   and we used the
properties 
\begin{mathletters}
\label{eqopcd}
\begin{eqnarray}
& & \hat C\hat C^\dagger = \hat C^\dagger\hat C = 1\\ & & \hat
C\,\sin{(\hat \omega_2 t)} =  \sin{(\hat \omega_1 t)}\;\hat C\\ & &
\hat C^\dagger\,\cos{(\hat \omega_1 t)} =  \cos{(\hat \omega_2
t)}\;\hat C^\dagger\\ & & \sqrt{\hat T\hat B_-}\,\hat\omega_2^n =
\hat\omega_1^n\,\sqrt{\hat T\hat B_-}\\ & & \sqrt{\hat B_+\hat
T^\dagger}\,\hat\omega_1^n =  \hat\omega_2^n\,\sqrt{\hat B_+\hat
T^\dagger}\,,
\end{eqnarray}
\end{mathletters}
proved in the appendix A of the Ref. \cite{ref17a},  together
with the operators relations 
\begin{mathletters}
\label{eqprch}
\begin{eqnarray}
& & \hat C\,\sqrt{\hat B_+\hat T^\dagger} = - \sqrt{\hat T\hat
B_-}\,\hat C^\dagger = i\hat H_2^{1/4}\\ & & \sqrt{\hat B_+\hat
T^\dagger}\,\hat C  = - \hat C^\dagger  \,\sqrt{\hat T\hat B_-}  =
i\hat H_1^{1/4}\,. 
\end{eqnarray}
\end{mathletters}

At this point, if we remember that \
$\left[\hat\nu_j,\hat\omega_j\right] = 0$,  $(j = 1,\, {\rm or}\; 2)$,
then we conclude that we can use the trigonometric relations involving
the product of trigonometric function with arguments $\hat\nu_jt$ and
$\hat\omega_jt$ because, in this case, we know that \
$\exp{(\hat\nu_jt)}\,\exp{(\pm\hat\omega_jt)} =
\exp{[(\hat\nu_j\pm\hat\omega_j)t]}$.  Now, using this fact, the
commutators 
\begin{equation}
\left[\hat\nu_1,\hat H_2\right] = \left[\hat\omega_1,\hat H_2\right] =
\left[\hat\nu_2,\hat H_1\right] = \left[\hat\omega_2,\hat H_1\right] =
0\,,
\label{eqcomm}
\end{equation}
and the properties (\ref{eqopcd}), we can show that 
\begin{mathletters}
\label{eqlyfele}
\begin{eqnarray}
\hat y_1(t)\,\hat F_{11}(t) &=&  i{\gamma\over 2}\sqrt{\hat T\hat
B_-}\left\{\cos{\left[(\hat\nu_2-\hat\omega_2)t\right]}\,
\sin{(\hat\omega_1t)} + \cos{\left[(\hat\nu_2+\hat\omega_2)t\right]}\,
\sin{(\hat\omega_1t)}\right\}\hat H_2^{1/4}\nonumber\\ &+&
i{\gamma\over 2}\hat
H_2^{1/4}\left\{\sin{\left[(\hat\nu_1-\hat\omega_1)t\right]}\,
\cos{(\hat\omega_2t)} - \sin{\left[(\hat\nu_1+\hat\omega_1)t\right]}\,
\cos{(\hat\omega_2t)} \right\}\sqrt{\hat B_+\hat T^\dagger}\\
\hat y_1(t)\,\hat F_{12}(t) &=&  {\gamma\over 2}\sqrt{\hat T\hat
B_-}\left\{\cos{\left[(\hat\nu_2-\hat\omega_2)t\right]}\,
\cos{(\hat\omega_1t)} + \cos{\left[(\hat\nu_2+\hat\omega_2)t\right]}\,
\cos{(\hat\omega_1t)}\right\}\sqrt{\hat T\hat B_-} \nonumber\\ &+&
{\gamma\over 2}\hat
H_2^{1/4}\left\{\sin{\left[(\hat\nu_1+\hat\omega_1)t\right]}\,
\sin{(\hat\omega_2t)} - \sin{\left[(\hat\nu_1-\hat\omega_1)t\right]}\,
\sin{(\hat\omega_2t)} \right\}\hat H_1^{1/4}\\
\hat y_2(t)\,\hat F_{21}(t) &=&  {\gamma\over 2}\sqrt{\hat B_+\hat
T^\dagger}\left\{\cos{\left[(\hat\nu_1-\hat\omega_1)t\right]}\,
\cos{(\hat\omega_2t)} + \cos{\left[(\hat\nu_1+\hat\omega_1)t\right]}\,
\cos{(\hat\omega_2t)}\right\}\sqrt{\hat B_+\hat T^\dagger} \nonumber\\
&+& {\gamma\over 2}\hat
H_1^{1/4}\left\{\sin{\left[(\hat\nu_2+\hat\omega_2)t\right]}\,
\sin{(\hat\omega_1t)} - \sin{\left[(\hat\nu_2-\hat\omega_2)t\right]}\,
\sin{(\hat\omega_1t)} \right\}\hat H_2^{1/4}\\
\hat y_2(t)\,\hat F_{22}(t) &=&  i{\gamma\over 2}\sqrt{\hat B_+\hat
T^\dagger}\left\{\cos{\left[(\hat\nu_1-\hat\omega_1)t\right]}\,
\sin{(\hat\omega_2t)} + \cos{\left[(\hat\nu_1+\hat\omega_1)t\right]}\,
\sin{(\hat\omega_2t)}\right\}\hat H_1^{1/4}\nonumber\\ &+&
i{\gamma\over 2}\hat
H_1^{1/4}\left\{\sin{\left[(\hat\nu_2-\hat\omega_2)t\right]}\,
\cos{(\hat\omega_1t)} - \sin{\left[(\hat\nu_2+\hat\omega_2)t\right]}\,
\cos{(\hat\omega_1t)} \right\}\sqrt{\hat T\hat B_-} \,. 
\end{eqnarray}
\end{mathletters}
In the same way, we can show that 
\begin{mathletters}
\label{eqlzfele}
\begin{eqnarray}
\hat z_1(t)\,\hat F_{11}(t) &=&  i{\gamma\over 2}\sqrt{\hat T\hat
B_-}\left\{\sin{\left[(\hat\nu_2-\hat\omega_2)t\right]}\,
\sin{(\hat\omega_1t)} + \sin{\left[(\hat\nu_2+\hat\omega_2)t\right]}\,
\sin{(\hat\omega_1t)}\right\}\hat H_2^{1/4}\nonumber\\ &-&
i{\gamma\over 2}\hat
H_2^{1/4}\left\{\cos{\left[(\hat\nu_1-\hat\omega_1)t\right]}\,
\cos{(\hat\omega_2t)} - \cos{\left[(\hat\nu_1+\hat\omega_1)t\right]}\,
\cos{(\hat\omega_2t)} \right\}\sqrt{\hat B_+\hat T^\dagger}\\
\hat z_1(t)\,\hat F_{12}(t) &=&  {\gamma\over 2}\sqrt{\hat T\hat
B_-}\left\{\sin{\left[(\hat\nu_2-\hat\omega_2)t\right]}\,
\cos{(\hat\omega_1t)} + \sin{\left[(\hat\nu_2+\hat\omega_2)t\right]}\,
\cos{(\hat\omega_1t)}\right\}\sqrt{\hat T\hat B_-} \nonumber\\ &-&
{\gamma\over 2}\hat
H_2^{1/4}\left\{\cos{\left[(\hat\nu_1+\hat\omega_1)t\right]}\,
\sin{(\hat\omega_2t)} - \cos{\left[(\hat\nu_1-\hat\omega_1)t\right]}\,
\sin{(\hat\omega_2t)} \right\}\hat H_1^{1/4}\\
\hat z_2(t)\,\hat F_{21}(t) &=&  {\gamma\over 2}\sqrt{\hat B_+\hat
T^\dagger}\left\{\sin{\left[(\hat\nu_1-\hat\omega_1)t\right]}\,
\cos{(\hat\omega_2t)} + \sin{\left[(\hat\nu_1+\hat\omega_1)t\right]}\,
\cos{(\hat\omega_2t)}\right\}\sqrt{\hat B_+\hat T^\dagger} \nonumber\\
&-& {\gamma\over 2}\hat
H_1^{1/4}\left\{\cos{\left[(\hat\nu_2+\hat\omega_2)t\right]}\,
\sin{(\hat\omega_1t)} - \cos{\left[(\hat\nu_2-\hat\omega_2)t\right]}\,
\sin{(\hat\omega_1t)} \right\}\hat H_2^{1/4}\\
\hat z_2(t)\,\hat F_{22}(t) &=&  i{\gamma\over 2}\sqrt{\hat B_+\hat
T^\dagger}\left\{\sin{\left[(\hat\nu_1-\hat\omega_1)t\right]}\,
\sin{(\hat\omega_2t)} + \sin{\left[(\hat\nu_1+\hat\omega_1)t\right]}\,
\sin{(\hat\omega_2t)}\right\}\hat H_1^{1/4}\nonumber\\ &-&
i{\gamma\over 2}\hat
H_1^{1/4}\left\{\cos{\left[(\hat\nu_2-\hat\omega_2)t\right]}\,
\cos{(\hat\omega_1t)} - \cos{\left[(\hat\nu_2+\hat\omega_2)t\right]}\,
\cos{(\hat\omega_1t)} \right\}\sqrt{\hat T\hat B_-} \,. 
\end{eqnarray}
\end{mathletters}

Now, the non-commutativity between the operators $\hat\omega_1$ and
$\hat\omega_2$ imply that to calculate the integrals involving the
terms given by the Eqs.~(\ref{eqlyfele}) and (\ref{eqlzfele}) we  need
to use the series expansion of the trigonometric functions.  In this
case the integrals can be easily done because the time  variable can
be considered as a parameter factor. Finally, using  these results
into Eq.~(\ref{eqvarp}) is trivial to find the expression
(\ref{eqs3p11}) for the matrix elements of the particular solution. 


\end{document}